\documentclass{PoS}

\title{ATLAS SUSY search prospects at 10 TeV}

\ShortTitle{ATLAS SUSY search prospects at 10 TeV}

\author{\speaker{J. Dietrich (for the ATLAS collaboration)}\thanks{I would like to thank the organisers of the
EPS09 for the invitation to present a poster and everyone in the ATLAS Collaboration whose work contributed to
this poster. The author acknowledges the support by the Landesstiftung Baden W\"urttemberg and the BMBF.}\\
        Physikalisches Institut, Albert-Ludwigs-Universit\"at Freiburg, Germany\\
        E-mail: \email{janet.dietrich@cern.ch}}

\abstract{The search for physics beyond the Standard Model (BSM) is one 
of the most important goals for the general purpose detector ATLAS at the Large Hadron Collider at CERN.
Already with early LHC data, the ATLAS experiment should be sensitive to discover physics beyond the Standard Model. 
This paper summarizes the prospects of the ATLAS experiment to find experimental evidence for Supersymmetry (SUSY) and Universal Extra Dimensions (UED) in channels with jets, leptons 
and missing transverse energy for an integrated luminosity of ${\cal L} = 200 \rm{pb}^{-1}$
at a centre-of-mass energy $\sqrt{s}$ = $10$~TeV. \\
Only a selection of the results is presented focussing on the the discovery reach for inclusive searches.}
\FullConference{The 2009 Europhysics Conference on High Energy Physics,\\
		 July 16 - 22 2009\\
		 Krakow, Poland}

\begin{document}
\section*{Inclusive searches for SUSY signals}
%\newtheorem{Inclusive}{Inclusive searches for SUSY signals}
%\begin{Inclusive}
\setlength{\parindent}{0cm}
%\paragraph{Inclusive searches for SUSY signals} 
%The search for physics beyond the Standard Model (BSM) is one 
%of the most important goals for the general purpose detector ATLAS at 
%the Large Hadron Collider at CERN. %After a short test run in 2008, the LHC, 
%%a proton-proton collider with a designed centre-of-mass energy of $\sqrt{s} =14$~TeV, will start in autumn 
%%2009 and provide an excellent opportunity to explore such new physics. 
%This paper summarises some strategies of the ATLAS experiment~\cite{ATLAS} to search for direct experimental
%evidence of Supersymmetry (SUSY) and Universal Extra Dimensions for an integrated luminosity of $200 \rm{pb}^{-1}$
%at a centre-of-mass energy $\sqrt{s}$ = $10$ TeV. Only a selection of the results is shown, focussing on the discovery reach for inclusive searches in channels with a combination of jets, leptons and missing transverse energy. 
%More  details can be found in ~\cite{atlas_pub_note}.\\ 
Searches for SUSY have to deal with models that have a relatively large set of free parameters. 
In this article I will focus on R-parity conserving SUSY particle production. 
%To enlarge previous studies~\cite{ATLAS} also models in the phenomenological minimal supersymmetric Standard Model (pMSSM) and 
In order to cover many different topologies both models in the phenomenological minimal supersymmetric Standard Model (pMSSM)~\cite{tom_rizzo} and 
SUSY look alike scenarios such as Universal Extra Dimensions (UED) with Kaluza-Klein-parity conservation were 
considered.\\ 
%For all these models the strongly interacting SUSY particles (squarks and/or gluinos) are produced in 
%pairs and are unstable. Each will decay via a complicated series of cascade processes 
%into states which include high $p_{T}$ Standard Model particles and the lightest SUSY particles, 
%that escapes the detector unseen. Therefore SUSY search strategies concentrate on events 
%with large missing transverse energy $E_{T}^{miss}$ and reconstructed particles with large 
%transverse momentum like jets or leptons, which number strongly depends on 
%the cascade decay of the squarks and/or gluinos. 
ATLAS studied various channels with different numbers of jets ($\ge 2$, $\ge 3$, $\ge 4$) and leptons (0,1,2),
trying to keep the SUSY searches robust and inclusive in order to cover as many signatures and topologies as 
possible. The applied selection cuts used in the inclusive SUSY searches are for example: require hadronic jets and $E_{T}^{miss}$ above a certain threshold, and spherical events. In the no-lepton search mode events with an isolated high $p_{T}$ ($>20$~GeV) electron
or muon are vetoed while for the 1 (2) lepton mode one (two) identified high $p_T$ lepton(s) is (are) required.
Additionally a cut on the transverse mass $M_{T}$, constructed from the identified lepton and 
the missing transverse energy, is applied in the one lepton channel. A full description of all selection cuts is presented in ~\cite{atlas_pub_note}. %All cuts on the number of jets 
%and on the transverse momentum of jets are common to the channels with different
%lepton multiplicities.\\
The final discriminating variable between SUSY and background we have used to search for an excess of events in 
various channels is $M_{\rm{eff}}$. It is defined as the scalar sum of transverse momenta 
of all main objects as: $M_{eff} \equiv \sum_{i=1}^{N_{jets}} P_{T}^{{\rm jet},i} + 
   \sum_{i=1}^{N_{lep}} P_{T}^{{\rm lep},i} + E_{T}^{miss}$
where $N_{jets}$ is the number of jets (2-4) and
$N_{lep}$ is the number of leptons (0-2). Further high $P_T$ jets or leptons are not included in the sum.

\section*{Discovery reach}
\begin{figure*}[htb]
\centering
\includegraphics[width=55mm]{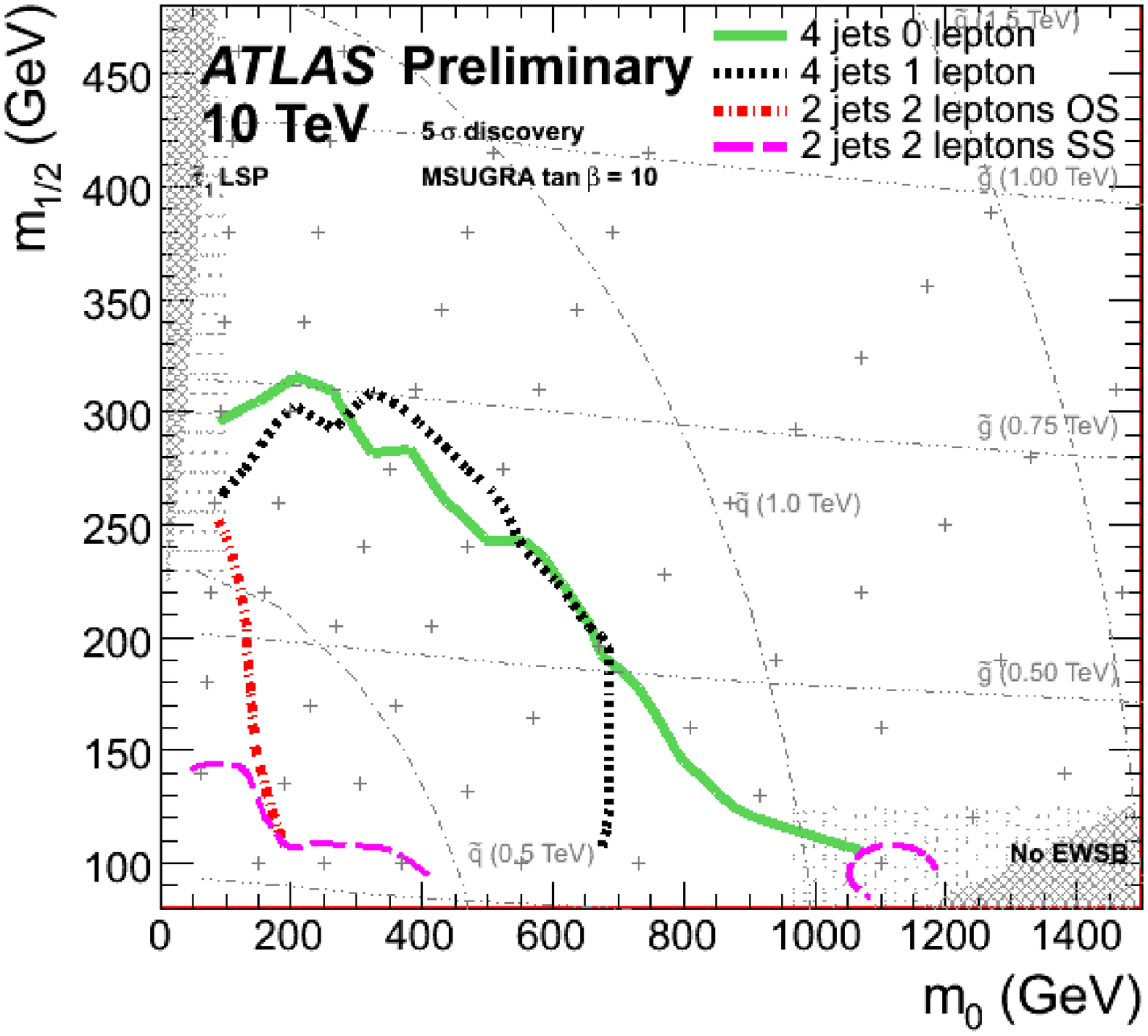}
\includegraphics[width=55mm]{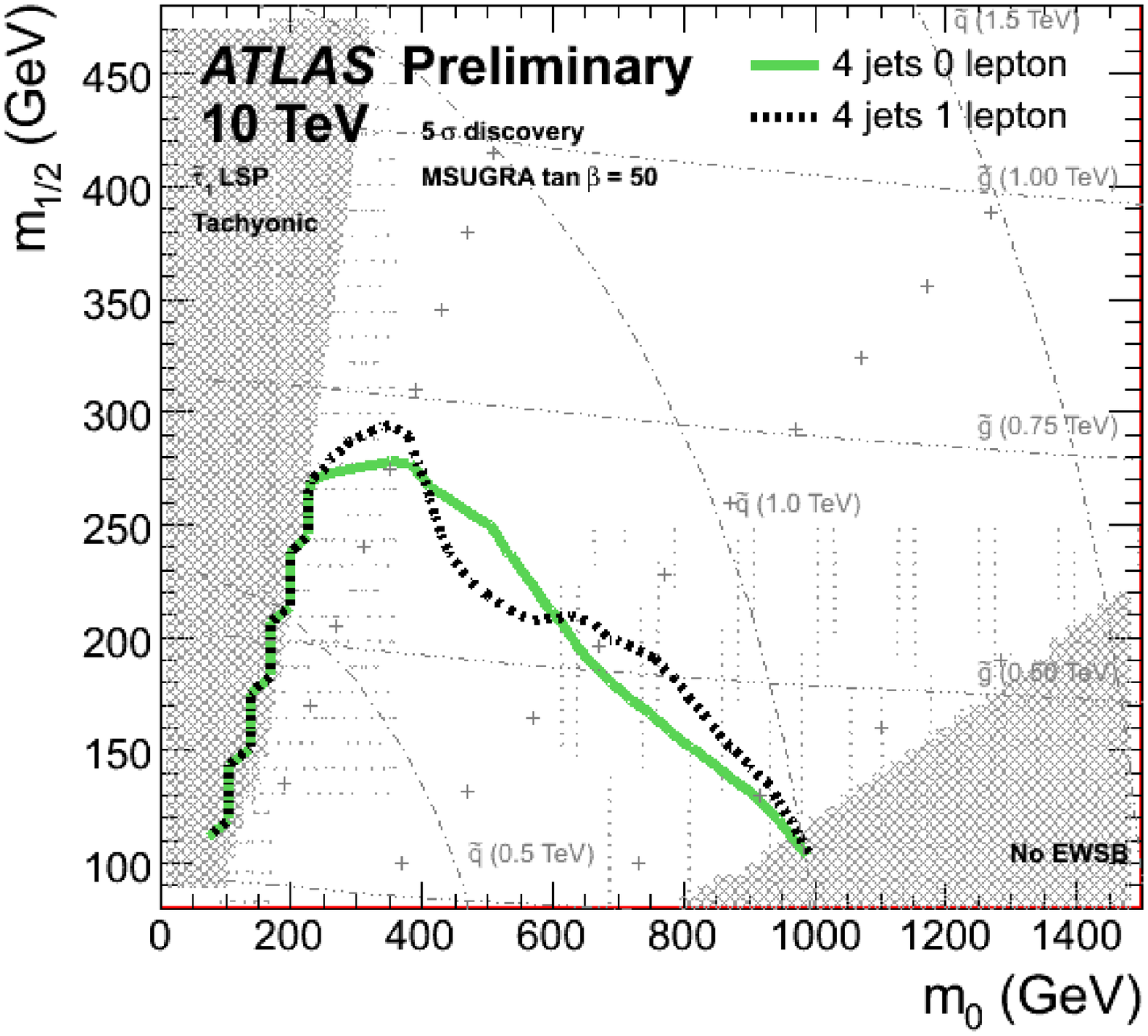}
\caption{The 5$\sigma$ contour lines for the ATLAS experiment for 200 $\rm{pb}^{-1}$ at $\sqrt{s} =10$~TeV for the 4 jet 0 lepton, the 4 jet 1 lepton and the 2 jet 2 lepton channel as a function of $M_0$ and $M_{1/2}$ for the mSUGRA model with $\tan \beta =10$ (left) and with $\tan \beta =50$ (right).}\label{reach}
\label{reach_msugra}
\end{figure*}

\setlength{\parindent}{0cm}
We have explored the reach of our search strategies by studying grids of models in the parameter space of mSUGRA, pMSSM and UED. For each point in these grids the same set of selection cuts are applied and the significance is calculated. Note that for the significance calculation
a systematic uncertainty of $50\%$ on the SM background estimate was taken into account. For a detailed explanation of the statistical procedure see \cite{ATLAS} page 1590-1591.\\
%The discovery reach plots are all made by finding the optimal $M_{eff}$ cut (in steps of 400 GeV) to maximize the significance $Z_N$. For a detailed explanation of the statistical procedure 
%see \cite{atlas} page 1590-1591. Note that for the significance calculation a uncertainty of $50\%$ was taken into account (for more details see \cite{atlas_pub_note}).
The following plots show only the channels with the largest discovery reach for each lepton 
multiplicity.  No attempt was made to combine the significance of the various channels. The 5$\sigma$ discovery reach lines in the $M_{0}$ - $M_{1/2}$ plane for different channels 
for the mSUGRA model with $A_0=0$~GeV, $\mu > 0$, $\tan \beta =10$ and $\tan \beta =50$, are shown in figure
\ref{reach_msugra}. The plots show only the 4jet 0 lepton, 4jet 1 lepton channel and 2 jet 2lepton channel with opposite charges (left plot). The 0 and 1 lepton channels have similar potential for a discovery in the studied mSUGRA grids. The discovery reach for the 2 leptons channel with same sign charges is taken from reference ~\cite{wisconsin}.
 
\begin{figure*}[htb]
\centering
\includegraphics[width=55mm]{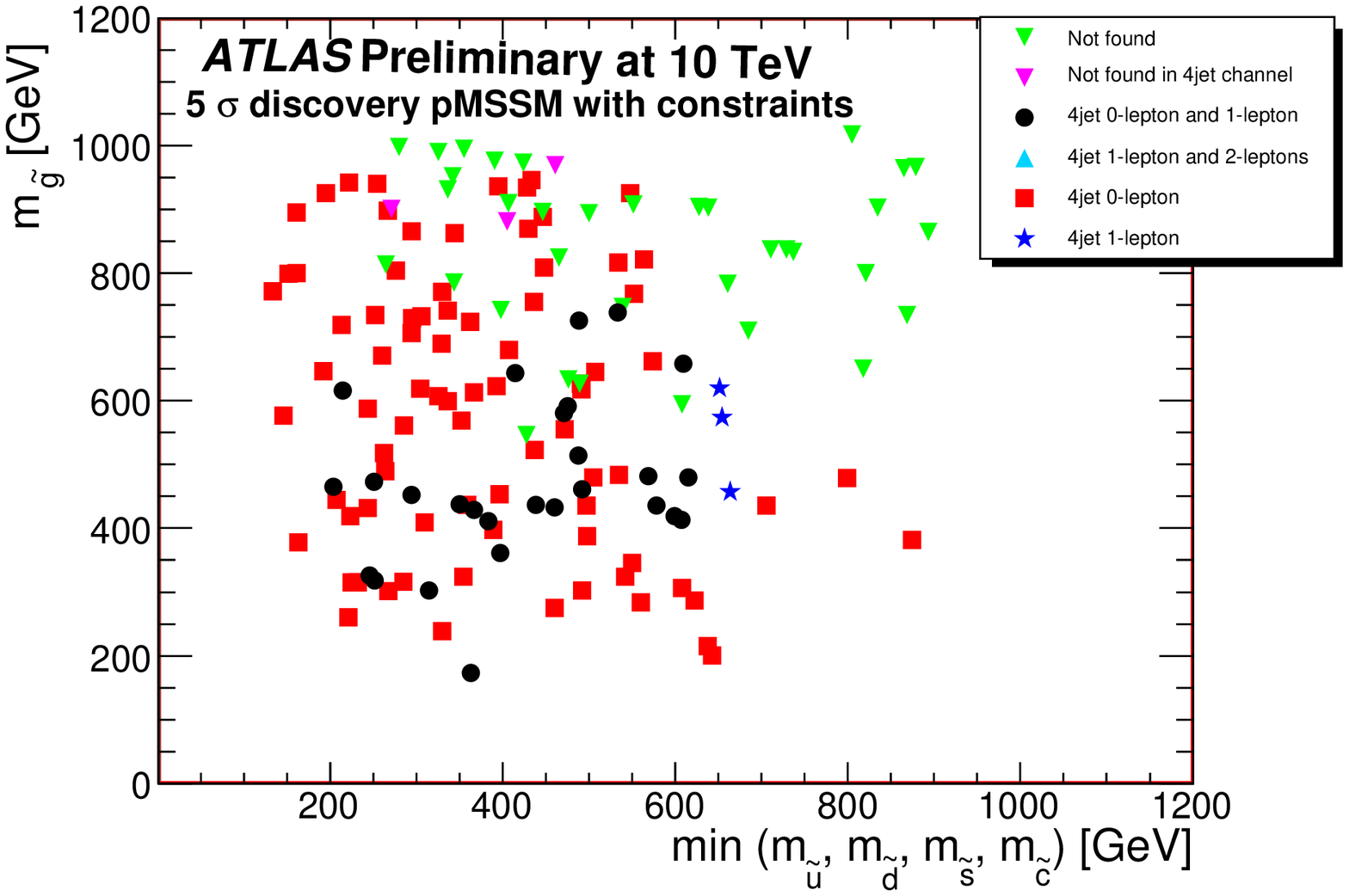}
\includegraphics[width=55mm]{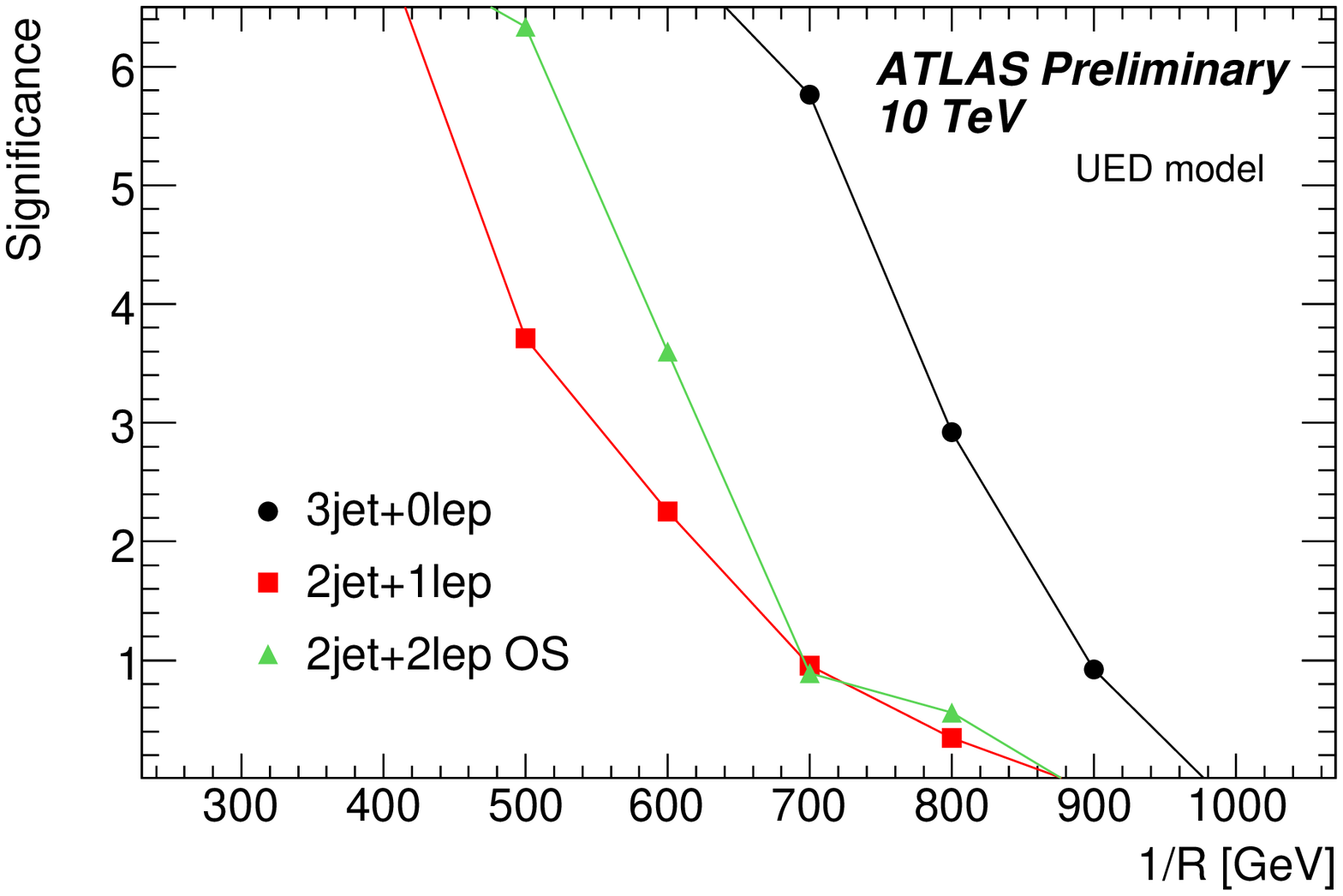}
\caption{The points of the pMSSM grids with constraints as a function of the minimal mass of the light squarks and the
gluino mass (left) and the significance $\sigma$ as a function of $1/R$ for the Universal Extra Dimensions scenario, taking
into account channels with 0, 1 and 2 leptons (right).}\label{reach}
\label{reach_MSSM}
\end{figure*}

\renewcommand{\thefootnote}{\fnsymbol{footnote}} 
Figure \ref{reach_MSSM} shows the 5$\sigma$ discovery reach for the pMSSM grid
with constraints as a function of the minimal mass of the first and second generation
squarks and the mass of the gluino (left side). Most considered SUSY signals can be discovered 
with the 4 jets channels if the cross section is larger than $10$~pb and for squark and gluino masses up to 600 GeV.
A few points are only found with the 2 or 3 jets channels with 0 or 1 lepton.
In general the 4jets 0-lepton channel is more effective than the 1-lepton one, because many points
do not lead to significant high $p_T$ lepton production. \\
The discovery reach for the UED model as a function of $1/R$ \footnote[2]{For UED models the extra dimension is compactified with the radius R that defines the size of the extra dimension.}, for the 3 jet 0-lepton channel, 
the 2 jet 1 lepton and the 2 jet channel with 2 leptons with opposite sign charges is shown in 
figure \ref{reach_MSSM} on the right side. The largest discovery reach is found in the 3 jet 0 lepton channel.
A $5\sigma$ significance can be achieved up to $1/R \approx 700$ ~GeV with this channel.

\section*{Conclusion}
\setlength{\parindent}{0cm}
The discovery potential for inclusive SUSY search channels with 0 leptons, 1 lepton or 2 opposite sign leptons and
$\ge 2$, $\ge 3$ or $\ge 4$ jets have been investigated for a scenario assuming an LHC centre-of-mass energy of 
$\sqrt{s}$ = $10$ TeV and an integrated luminosity of ${\cal L} = 200 \rm{pb}^{-1}$.
The results of the scans show that ATLAS could discover
signals of R-parity conserving SUSY with gluino and squark masses less than 600-700 GeV in many scenarios.
Signals of Universal Extra Dimensions can be discovered if $1/R < 700$~GeV.

%\section{Acknowledgments}
%I would like to thank the organisers of the EPS09 for the invitation to present a poster and everyone in the ATLAS Collaboration whose work contributed to this poster.

\end{document}